\documentstyle[12pt]{article}        
\newlength{\dinwidth}                       
\newlength{\dinmargin}                      
\def\lsim{\mathrel{\rlap{\lower4pt\hbox{\hskip1pt$\sim$}}
    \raise1pt\hbox{$<$}}}                
\def\gsim{\mathrel{\rlap{\lower4pt\hbox{\hskip1pt$\sim$}}
    \raise1pt\hbox{$>$}}}                
\input{epsfig.sty}

\newcommand{\be}{\begin{eqnarray}}
\newcommand{\ee}{\end{eqnarray}}

\begin{document}

\Huge{\noindent{Istituto\\Nazionale\\Fisica\\Nucleare}}

\vspace{-3.9cm}

\Large{\rightline{Sezione SANIT\`{A}}}
\normalsize{}
\rightline{Istituto Superiore di Sanit\`{a}}
\rightline{Viale Regina Elena 299}
\rightline{I-00161 Roma, Italy}

\vspace{0.65cm}

\rightline{INFN-ISS 96/5}
\rightline{June 1996}

\vspace{2cm}

\begin{center}

\Large{INVESTIGATION OF THE NEUTRON STRUCTURE FUNCTION VIA
SEMI-INCLUSIVE DEEP INELASTIC ELECTRON SCATTERING OFF THE DEUTERON}
\footnote{To appear in the Proceedings of the Workshop on {\em Future
Physics at HERA}, DESY (Germany), September 1995 to May 1996.}\\

\vspace{1cm}

\large{Silvano Simula}\\

\vspace{0.5cm}

\normalsize{Istituto Nazionale di Fisica Nucleare, Sezione Sanit\`{a},\\
Viale Regina Elena 299, I-00161 Roma, Italy}

\end{center}

\vspace{1cm}

\begin{abstract}

The production of slow nucleons in semi-inclusive deep inelastic electron
scattering off the deuteron is investigated in the region $x \gsim 0.3$
for kinematical conditions accessible at $HERA$. Within the spectator
mechanism the semi-inclusive cross section exhibits a scaling property,
which can be used as a model-independent test of the dominance of the
spectator mechanism itself, providing in this way an interesting tool to
investigate the neutron structure function. The possibility of extracting
model-independent information on the neutron to proton structure function
ratio from semi-inclusive experiments is also illustrated.

\end{abstract}

\newpage

\vspace*{10cm}

\newpage

\pagestyle{plain}

\vspace*{1cm}
\begin{center}  \begin{Large} \begin{bf}
Investigation of the neutron structure function via\\ semi-inclusive deep
inelastic electron scattering\\ off the deuteron\\
  \end{bf}  \end{Large}
  \vspace*{5mm}
  \begin{large}
Silvano SIMULA\\ 
  \end{large}
\end{center}
~~INFN, Sezione Sanit\`{a}, Viale Regina Elena 299, I-00161 Roma, Italy\\
\begin{quotation}
\noindent
{\bf Abstract:}
The production of slow nucleons in semi-inclusive deep inelastic electron
scattering off the deuteron is investigated in the region $x \gsim 0.3$ for
kinematical conditions accessible at $HERA$. Within the spectator mechanism the
semi-inclusive cross section exhibits a scaling property, which can be used as
a model-independent test of the dominance of the spectator mechanism itself,
providing in this way an interesting tool to investigate the neutron structure
function. The possibility of extracting model-independent information on the
neutron to proton structure function ratio from semi-inclusive experiments is
also illustrated.
\end{quotation}

\vspace{0.5cm}

\noindent Untill now, experimental information on the structure function of the
neutron has been inferred from nuclear (usually deuteron) deep inelastic
scattering ($DIS$) data \cite{DATA} by unfolding the neutron contribution from
the inclusive nuclear cross section. Such a procedure typically involves the
subtraction of both Fermi motion effects and contributions from different
nuclear constituents (i.e., nucleons, mesons, isobars, ...), leading to
non-trivial ambiguities related to the choice of the model used to describe the
structure of the target and the mechanism of the reaction. An interesting way
to get information on the neutron structure function could be the investigation
of semi-inclusive $DIS$ reactions of leptons off the deuteron. In Ref.
\cite{SIM96} the process  $^2H(\ell, \ell' N)X$ has been investigated at
moderate and large values of the Bjorken variable $x \equiv Q^2 / 2M \nu$ ($x
\gsim 0.3$) within the so-called spectator mechanism, according to which, after
lepton interaction with a quark of a nucleon in the deuteron, the spectator
nucleon is emitted because of recoil and detected in coincidence with the
scattered lepton. It has been shown \cite{SIM96} that the semi-inclusive cross
section corresponding to such a mechanism exhibits a scaling property (the
spectator scaling), which can be used as a model-independent test of the
dominance of the spectator mechanism itself. In the spectator-scaling regime
the neutron structure function can be investigated from semi-inclusive data
and, moreover, the neutron to proton structure function ratio $R^{(n/p)}(x,
Q^2) \equiv F_2^n(x, Q^2) / F_2^p(x, Q^2)$ can be obtained directly from the
ratio of the semi-inclusive cross sections of the processes $^2H(e,e'p)X$ and
$^2H(e,e'n)X$.

\indent The aim of this contribution is to address the issue of the spectator
scaling in case of electron kinematical conditions accessible at $HERA$. To
this end, let us briefly remind that the semi-inclusive cross section of the
process $^2H(e,e'N)X$ reads as follows
 \be
    {d^4 \sigma \over dE_{e'} ~ d\Omega_{e'} ~ dE_2 ~ d\Omega_2} = 
    \sigma_{Mott} ~ p_2 ~ E_2 ~ \sum_i ~ V_i ~ W_i^D(x, Q^2; \vec{p}_2)
     \label{1} 
 \ee
where $E_e$ ($E_{e'}$) is the initial (final) energy of the electron; $Q^2
\equiv - q^2 = |\vec{q}|^2 - \nu^2$ is the squared four-momentum transfer; $i
\equiv \{L, T, LT, TT\}$ identifies the different types of semi-inclusive
response functions ($W_i^D$) of the deuteron; $\vec{p}_2$ is the momentum of
the detected nucleon; $E_2 = \sqrt{M^2 + p_2^2}$ its energy ($p_2 \equiv
|\vec{p}_2|$) and $V_i$ is a virtual photon flux factor (see \cite{SIM96}).
Within the spectator mechanism the virtual photon is absorbed by a quark
belonging to the nucleon $N_1$ in the deuteron and the recoiling nucleon $N_2$
is emitted and detected in coincidence with the scattered electron. Thus, Eq.
(\ref{1}) can be written in terms of the structure function $F_2^{N_1}(x^*,
Q^2)$ of the struck nucleon as \cite{SIM96}
 \be
    {d^4 \sigma \over dE_{e'} ~ d\Omega_{e'} ~ dE_2 ~ d\Omega_2} = K ~ M ~ p_2 ~
    n^{(D)}(p_2) ~ {F_2^{N_1}(x^*, Q^2) \over x^*} ~ D^{N_1}(x, Q^2; \vec{p}_2)
    \label{8}
 \ee
where $n^{(D)}$ is the (non-relativistic) nucleon momentum distribution in the
deuteron, $x^* \equiv Q^2 / (Q^2 + M_1^{*2} - M^2)$ and $M_1^*$ is the invariant
mass of the struck nucleon, given through the energy and momentum conservations
by: $M_1^* = \sqrt{(\nu + M_D - E_2)^2 - (\vec{q} - \vec{p}_2)^2}$, with $M_D$
being the deuteron mass. In Eq. (\ref{8}) $K \equiv (2 M x^2 E_e E_{e'} / \pi
Q^2) ~ (4 \pi \alpha^2 / Q^4) ~ \left [ 1 - y + (y^2 / 2) + (Q^2 / 4 E_e^2)
\right ]$, with $y \equiv \nu / E_e$, and the quantity $D^{N_1}(x, Q^2;
\vec{p}_2)$ depends both upon kinematical factors and the ratio $R_{L/T}^{N_1}$
of the longitudinal to transverse cross section off the nucleon (see
\cite{SIM96}). The relevant quantity, which will be discussed hereafter, is
related to the semi-inclusive cross section (\ref{1}) by
 \be
     F^{(s.i.)}(x, Q^2; \vec{p}_2) & \equiv & {1 \over \tilde{K}} ~ {d^4 \sigma
     \over dE_{e'} ~ d\Omega_{e'} ~ dE_2 ~ d\Omega_2} \nonumber \\
     & = & M ~ p_2 ~ n^{(D)}(p_2) ~ {F_2^{N_1}(x^*, Q^2) \over x^*} ~
     \tilde{D}^{N_1}(x, Q^2; \vec{p}_2)
     \label{9}
 \ee
where $\tilde{K}$ is a kinematical factor given by $\tilde{K} \equiv K ~
[D^{N_1}(x, Q^2; \vec{p}_2)]_{R_{L/T}^{N_1} = 0}$ and $\tilde{D}^{N_1}(x, Q^2;
\vec{p}_2) =$ $D^{N_1}(x, Q^2; \vec{p}_2) ~ / ~ [D^{N_1}(x, Q^2;
\vec{p}_2)]_{R_{L/T}^{N_1} = 0}$. Eq. (\ref{9}) differs from the definition of
$F^{(s.i.)}$ given in \cite{SIM96}, for $\tilde{K}$, which incorporates
kinematical factors depending on $\vec{p}_2$, is used instead of $K$, which
depends only on the electron kinematical variables.

\begin{figure}[t]

\epsfig{file=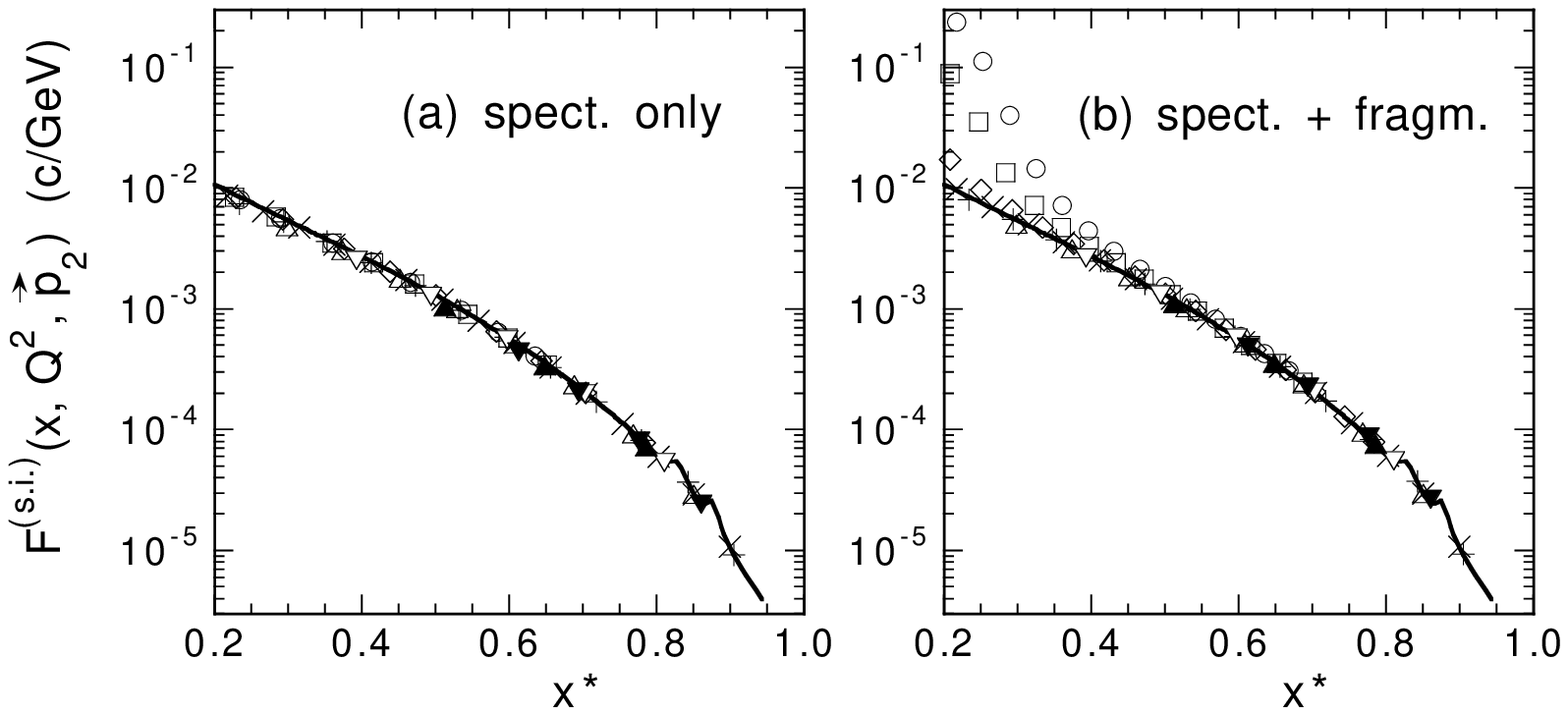}

\vspace{-21.0cm}

\parbox{0.25cm} \ $~$\ \parbox{16cm}{\small {\noindent \it Fig. 1. (a) The
function $F^{(s.i.)}(x, Q^2, \vec{p}_2)$ (Eq. (\ref{9})) for the process
$^2H(e,e'p)X$ plotted versus the spectator-scaling variable $x^*$ (Eq.
(\ref{12})) at $Q^2 = 10 ~ (GeV/c)^2$ and $p_2 = 0.5 ~ GeV/c$. The values of
$x$ have been varied in the range $0.20 \div 0.95$, whereas the various markers
correspond to different values of the nucleon detection angle $\theta_2$ chosen
in the range $10^o \div 150^o$. The solid line is the spectator-scaling
function $F^{(sp)}(x^*, Q^2, p_2)$ (Eq. (\ref{11})) calculated using the
deuteron momentum distribution corresponding to the Paris nucleon-nucleon
interaction \cite{PARIS} and to the parametrization of the neutron structure
function of Ref. \cite{SLAC}. (b) The same as in (a), but including the effects
of the target fragmentation of the struck nucleon, evaluated as in Ref.
\cite{SIM93}, as well as the contribution of the proton emission arising from
virtual photon absorption on $6q$ cluster configurations in the deuteron,
evaluated following Ref. \cite{SIM95} and adopting a $6q$ bag probability equal
to $2 \%$. In particular, the open dots and squares, which exhibit large
violations of the spectator scaling, correspond to $\theta_2 = 10^o$ and
$30^o$, respectively.}}

\end{figure}

\indent In the Bjorken limit one would expect $R_{L/T}^{N_1} \to_{Bj} 0$, so
that $\tilde{D}^{N_1} \to_{Bj} 1$, which implies $F^{(s.i.)}(x, Q^2; \vec{p}_2)
\to_{Bj} M ~ p_2 ~ n^{(D)}(p_2) ~ F_2^{N_1}(x^*) / x^*$, where $F_2^{N_1}(x^*)$
stands for the nucleon structure function in the Bjorken limit (apart from
logarithmic $QCD$ corrections). Therefore, in the Bjorken limit and at fixed
values of $p_2$ the function (\ref{9}) does not depend separately upon $x$ and
the nucleon detection angle $\theta_2$, but only upon the variable $x^*
\to_{Bj} x / (2 - z_2)$, with $z_2 = [E_2 - p_2 cos(\theta_2)] / M$ being the
light-cone momentum fraction of the detected nucleon. In what follows, we will
refer to the function $F^{(sp)}(x^*, Q^2, p_2)$ and variable $x^*$, given
explicitly by
 \be
    F^{(sp)}(x^*, Q^2, p_2) & \equiv & M ~ p_2 ~ n^{(D)}(p_2) ~ F_2^{N_1}(x^*,
    Q^2) ~ / ~ x^*
    \label{11}  \\
    x^* & \equiv & {Q^2 \over Q^2 + (\nu + M_D - E_2)^2 - (\vec{q} -
    \vec{p}_2)^2 - M^2}
    \label{12}
 \ee
as the spectator-scaling function and variable, respectively. The essence of
the spectator scaling relies on the fact that the variable $x^*$ gathers
different electron and nucleon kinematical conditions (in $x$ and $\theta_2$),
corresponding to the same value of the invariant mass produced on the struck
nucleon. The deuteron response will be the same only if the spectator
mechanism dominates and, therefore, the experimental observation of the
spectator scaling represents a test of the dominance of the spectator mechanism
itself.

\indent Eq. (\ref{9}) has been calculated considering electron kinematical
conditions accessible at $HERA$ (i.e., $E_e = 30 ~ GeV$ and $Q^2 = 10 ~
(GeV/c)^2$) and $p_2 = 0.1, ~ 0.3, ~ 0.5 ~ GeV/c$. The Bjorken variable $x$ and
the nucleon detection angle $\theta_2$ have been varied in the range $0.20 \div
0.95$ and $10^o \div 150^o$, respectively (for sake of simplicity, the polar
angle $\phi_2$ has been chosen equal to $0$). As for the nucleon structure
function, the parametrization of the $SLAC$ data of Ref. \cite{SLAC}, containing
$R_{L/T}^{N_1} \simeq 0.18$, has been adopted. The results of the calculations,
performed at $p_2 = 0.5 ~ GeV/c$, are shown in Fig. 1(a) and compared with the
spectator-scaling function (\ref{11}). It can clearly be seen that at $Q^2 = 10
~ (GeV/c)^2$ the spectator scaling is almost completely fulfilled (within $10
\%$) in the whole $x$-range, including the region at $x^* \gsim 0.8$, where
(small) contributions from nucleon resonances are still visible. Using the new
definition (\ref{9}), the spectator scaling is fulfilled not only at $ p_2
\lsim 0.3 ~ GeV/c$, as shown in \cite{SIM96}, but also at higher values of
$p_2$.

\indent In the spectator-scaling regime the measurement of the semi-inclusive
cross section both for the $^2H(e,e'p)X$ and $^2H(e,e'n)X$ processes would allow
the investigation of two spectator-scaling functions, involving the same
nuclear part, $M p_2 n^{(D)}(p_2)$, and the neutron and proton structure
functions, respectively. Therefore, assuming $R_{L/T}^n = R_{L/T}^p$ (as it is
suggested by recent $SLAC$ data analyses \cite{RLT_SLAC}), both the nuclear part
and the factor $\tilde{D}^{N_1}(x, Q^2; \vec{p}_2)$ cancel out in the ratio
$R^{(s.i.)}(x, Q^2, \vec{p}_2) \equiv d^4 \sigma[^2H(e,e'p)X] / d^4
\sigma[^2H(e,e'n)X]$, which provides in this way directly the neutron to proton
structure function ratio $R^{(n/p)}(x^*, Q^2)$. With respect to the function
$F^{(s.i.)}(x, Q^2, \vec{p}_2)$, the ratio $R^{(s.i.)}(x, Q^2, \vec{p}_2)$
exhibits a more general scaling property, for at fixed $Q^2$ it does not depend
separately upon $x$, $p_2$ and $\theta_2$, but only on $x^*$. This means that
any $p_2$-dependence of the ratio $R^{(s.i.)}$ would allow to investigate
off-shell deformations of the nucleon structure functions (see below).

\indent The results presented and, in particular, the spectator-scaling
properties of $F^{(s.i.)}$ and $R^{(s.i.)}$ could in principle be modified by
the effects of mechanisms different from the spectator one, like, e.g., the
fragmentation of the struck nucleon, or by the breakdown of the impulse
approximation. In order to estimate the effects of the so-called target
fragmentation of the struck nucleon (which is thought to be responsible for the
production of slow hadrons in $DIS$ processes), we adopt the approach of Ref.
\cite{SIM93}, where the hadronization mechanism has been parametrized through
the use of fragmentation functions, whose explicit form has been chosen
according to the prescription of Ref. \cite{BDS92}, elaborated to describe the
production of slow protons in $DIS$ of (anti)neutrinos off hydrogen and
deuterium targets. Furthermore, the effects arising from possible six-quark
($6q$) cluster configurations at short internucleon separations, are explicitly
considered. According to the mechanism first proposed in Ref.
\cite{SIX-QUARKS}, after lepton interaction with a quark belonging to a $6q$
cluster, nucleons can be formed out of the penta-quark residuum and emitted
forward as well as backward. The details of the calculations can be easily
inferred from Ref. \cite{SIM95}, where $6q$ bag effects in semi-inclusive $DIS$
of leptons off light and complex nuclei have been investigated. The estimate of
the nucleon production, arising from the above-mentioned target fragmentation
processes, is shown in Fig. 1(b) for the function $F^{(s.i.)}$ and in Fig. 2
for the ratio $R^{(s.i.)}$. It can clearly be seen that: i) only at $x^* \lsim
0.4$ the fragmentation processes can produce relevant violations of the
spectator scaling (see Figs. 1(b) and 2(a)); ii) backward kinematics (see Fig.
2(b)) appear to be the most appropriate conditions to extract the neutron to
proton ratio $R^{(n/p)}$. Moreover, explicit calculations show that the
relevance of the fragmentation processes drastically decreases when $p_2 < 0.5
~ GeV/c$.

\begin{figure}[t]

\epsfig{file=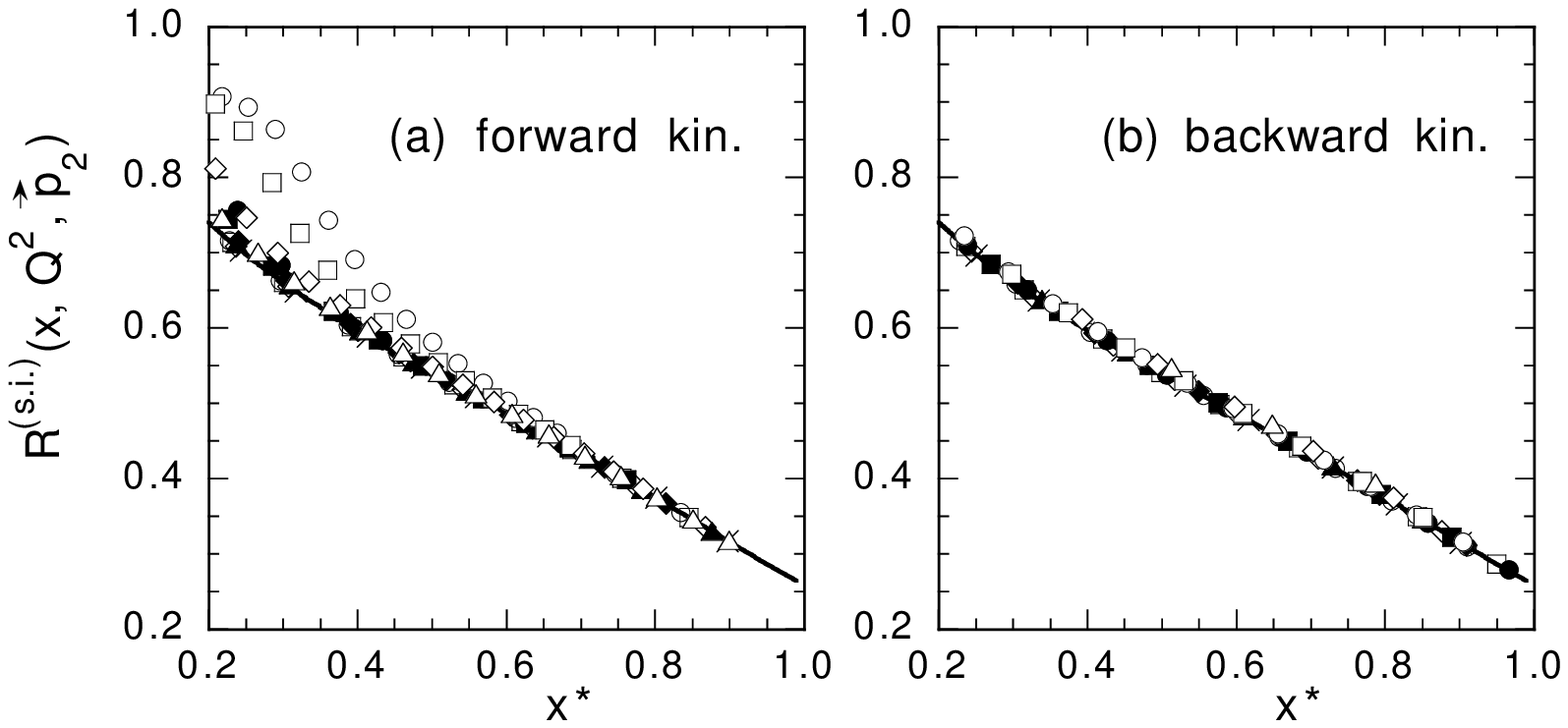}

\vspace{-21.0cm}

\parbox{0.25cm} \ $~$\ \parbox{16cm}{\small {\noindent \it Fig. 2. The ratio
$R^{(s.i.)}(x, Q^2, \vec{p}_2)$ of the semi-inclusive cross sections for the
processes $^2H(e,e'p)X$ and $^2H(e,e'n)X$, calculated at $Q^2 = 10 ~ (GeV/c)^2$
and $p_2 = 0.1, 0.3, 0.5 ~ GeV/c$. The values of $x$ have been varied in the
range $0.20 \div 0.95$. Forward ($\theta_2 < 90^o$) and backward ($\theta_2 >
90^o$) nucleon emissions are shown in (a) and (b), respectively. The solid line
is the neutron to proton structure function ratio $R^{(n/p)}(x^*, Q^2)$
calculated using the parametrization of the nucleon structure function of Ref.
\cite{SLAC}.}}

\end{figure}

\indent As far as the impulse approximation is concerned, it should be reminded
that our calculations have been performed within the assumption that the debris
produced by the fragmentation of the struck nucleon does not interact with the
recoiling spectator nucleon. Estimates of the final state interactions of the
fragments in semi-inclusive processes off the deuteron have been obtained in
\cite{TN92}, suggesting that rescattering effects should play a minor role
thanks to the finite formation time of the dressed hadrons. Moreover, backward
nucleon emission is not expected to be sensitively affected by forward-produced
hadrons (see \cite{BDT94}), and final state interaction effects are expected to
cancel out (at least partially) in the cross section ratio $R^{(s.i.)}(x, Q^2,
\vec{p}_2)$. Besides fragmentation processes and final state interactions, also
nucleon off-shell effects \cite{OFF-SHELL} might produce violations of the
spectator scaling, in particular at high values of $p_2$ ($\gsim 0.3 ~ GeV/c$).
The results of the calculations of the ratio $R^{(s.i.)}(x, Q^2, \vec{p}_2)$,
obtained considering the off-shell effects suggested in Refs. \cite{DT86} and
\cite{HT90}, are shown in Fig. 3(a) and 3(b), respectively. It can be seen that
the measurement of the ratio $R^{(s.i.)}(x, Q^2, \vec{p}_2)$ represents an
interesting tool both to investigate the ratio of free neutron to proton
structure function, provided $p_2 \sim 0.1 \div 0.2 ~ GeV/c$, and to get
information on the possible off-shell behaviour of the nucleon structure
function when $p_2 \gsim 0.3 ~ GeV/c$.

\begin{figure}[t]

\epsfig{file=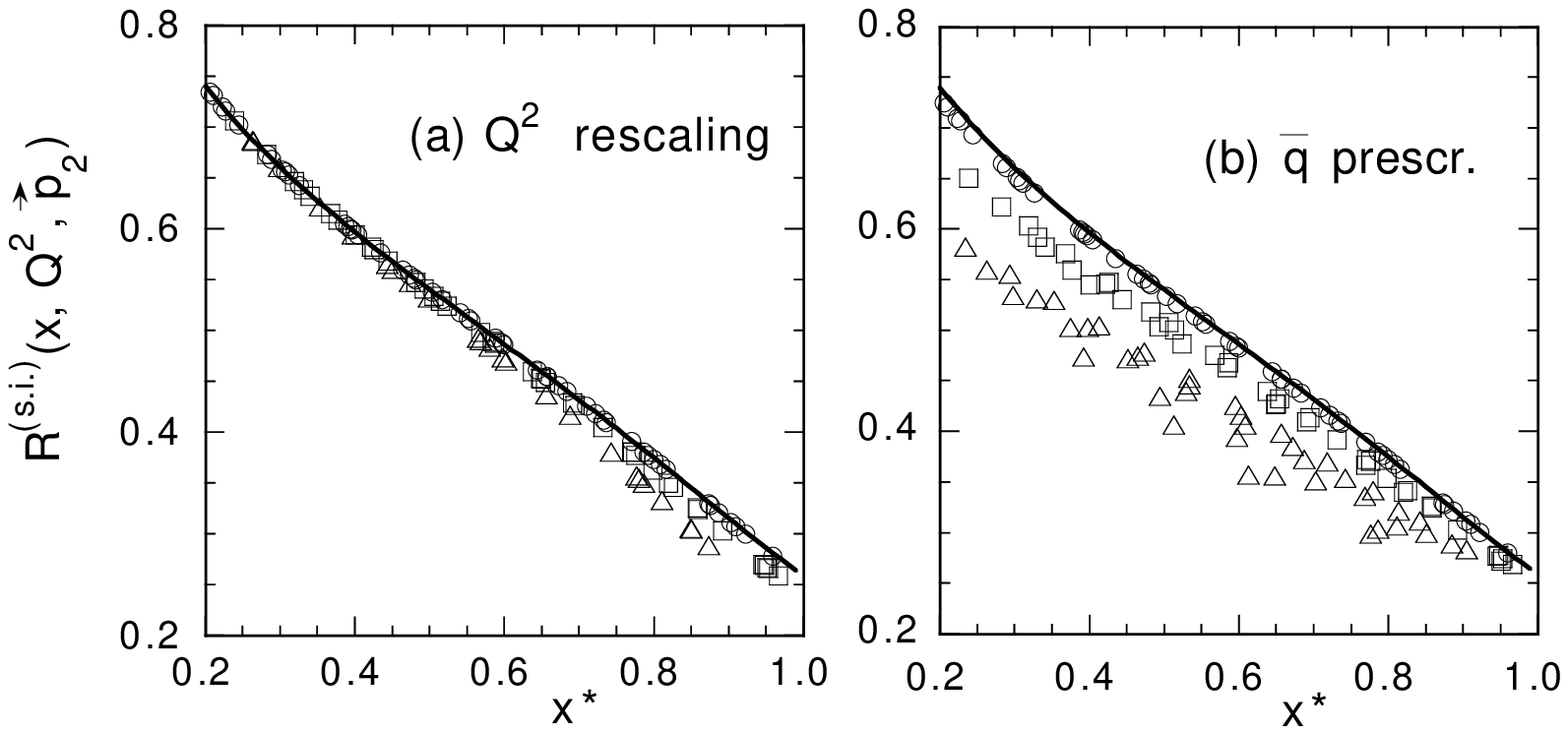}

\vspace{-21.0cm}

\parbox{0.25cm} \ $~$\ \parbox{16cm}{\small {\noindent \it Fig. 3. The ratio
$R^{(s.i.)}(x, Q^2, \vec{p}_2)$ of the semi-inclusive cross sections for the
processes $^2H(e,e'p)X$ and $^2H(e,e'n)X$, calculated considering the off-shell
effects proposed in Refs. \cite{DT86} (a) and \cite{HT90} (b), respectively.
Backward nucleon kinematics only ($\theta_2 > 90^o$) have been considered. The
dots, squares and triangles correspond to $p_2 = 0.1, ~ 0.3, ~ 0.5 ~ GeV/c$,
respectively. The solid line is the same as in Fig. 2.}}

\end{figure}

\indent In conclusion, the production of slow nucleons in semi-inclusive deep
inelastic electron scattering off the deuteron has been investigated in
kinematical regions accessible at $HERA$. Within the spectator mechanism the
semi-inclusive cross section exhibits the spectator-scaling property, which can
be used as a model-independent test of the dominance of the spectator mechanism
itself. In the spectator-scaling regime both the neutron structure function and
the neutron to proton structure function ratio can be obtained directly from
semi-inclusive $DIS$ data off the deuteron. Finally, the pattern of possible
spectator-scaling violations could provide relevant information on the off-shell
behaviour of the nucleon structure function in the medium.

\end{document}